\documentclass[12pt]{article}
\usepackage{amsmath}
\def\ex{{\hbox{\rm e}}}
 \def\mod{{\hbox{\rm mod}}}
\def\tr{{\hbox{\rm Tr}}}

\def\bea{\begin{eqnarray}}
\def\eea{\end{eqnarray}}
\def\beal{\begin{align}}
\def\eeal{\end{align}}
\def\ret{\nonumber\\{}}
\def\la#1{\label{#1}}
\def\eqs#1{(\ref{#1})}

\def\to{\rightarrow}

\def\sqr#1#2{{\vcenter{\vbox{\hrule height.#2pt
  \hbox{\vrule width.#2pt height#1pt \kern#1pt
    \vrule width.#2pt}
  \hrule height.#2pt}}}}

\def\raiz{\sqrt{2}}

\def\dalpha{{\dot\alpha}}

\def\mapright#1{\smash{\mathop{\longrightarrow}\limits^{#1}}}

\font\upright=cmu10 
\font\cmss=cmss10 at 12pt \font\cmsss=cmss8 at 8pt
\font\cmssi=cmss10 at 10pt

\def\IZ{\relax\ifmmode\mathchoice
{\hbox{\cmss Z\kern-.4em Z}}{\hbox{\cmss Z\kern-.4em Z}}
{\lower.4pt\hbox{\cmsss Z\kern-.4em Z}}
{\lower1.2pt\hbox{\cmsss Z\kern-.4em Z}}\else{\cmss Z\kern-.4em
Z}\fi}
\def\iz{\relax\hbox{\cmssi Z\kern -.4em Z}}
\def\mt{\rlap{\cmss T}\kern 3.0pt{\hbox{{\cmss T}}}}
\def\identity{{\upright\rlap{1}\kern 2.0pt 1}}
\def\inbar{\vrule height1.5ex width.4pt depth0pt}
\def\mininbar{\vrule height.75ex width.3pt depth0pt} 
\def\cc{\relax\,\hbox{$\mininbar\kern-.2em{\hbox{\rm\tiny
C}}$}}
\def\IC{\relax\,\hbox{$\inbar\kern-.3em{\rm C}$}}
\def\IR{\relax{\rm I\kern-.18em R}}

\def\IL{\relax{\rm I\kern-.18em L}}
\def\IH{\relax{\rm I\kern-.18em H}}
\def\IB{\relax{\rm I\kern-.18em B}}
\def\ID{\relax{\rm I\kern-.18em D}}
\def\IE{\relax{\rm I\kern-.18em E}}
\def\IF{\relax{\rm I\kern-.18em F}}
\def\IG{\relax\hbox{$\inbar\kern-.3em{\rm G}$}}
\def\IGa{\relax\hbox{${\rm I}\kern-.18em\Gamma$}}
\def\IH{\relax{\rm I\kern-.18em H}}
\def\II{\relax{\rm I\kern-.18em I}}
\def\IK{\relax{\rm I\kern-.18em K}}
\def\IP{\relax{\rm I\kern-.18em P}}
\def\IQ{\relax\hbox{$\inbar\kern-.3em{\rm Q}$}}
\def\cp{\IC\IP}
\def\hat{\widehat}
\def\C#1{{\cal #1}}
\def\cn{{\cal N}}

\begin{document}
\begin{titlepage}

\begin{flushright}   US-FT-6/99\\ hep-th/9903172\\ 
\end{flushright}
\vspace*{20pt}
\bigskip
\begin{center} {\huge The Vafa-Witten Theory for}
\vskip3mm {\huge  Gauge Group $SU(N)$}
\vskip 0.9truecm

\vspace{3pc}

{\large{J. M. F. Labastida and  Carlos Lozano}\footnote{e-mail: 
lozano@fpaxp1.usc.es}}

\vspace{1pc}

{\em  Departamento de F\'\i sica de Part\'\i culas,\\
Universidade de Santiago de Compostela,\\ E-15706 Santiago de
Compostela, Spain.\\}

\vspace{6pc}
\hrulefill

\vspace{.5pc}
{\large \bf Abstract}
\end{center} 
\noindent We derive the partition function for the Vafa-Witten
twist of  the $\cn=4$ supersymmetric gauge theory  with
gauge group $SU(N)$ (for prime $N$) and arbitrary values of the
't Hooft fluxes $v\in H^{2}(X,\IZ_{N})$ on K\"ahler
four-manifolds with $b^{+}_2>1$. 

\vspace{3pc}

\vspace{-.5pc}
\vspace{-.5pc}

\noindent\hrulefill

\vspace{5pc}
\noindent\today
\end{titlepage}

\def\theequation{\thesection.\arabic{equation}}

\section{Introduction}
\setcounter{equation}{0}

The study of topological quantum field theories (TQFTs) originated from the
twist of $\cn=4$ supersymmetric Yang-Mills theory has been pursued during
the last few years. These studies have led to the full solution for some of
the models involved, and have provided important tests of our ideas on
duality for Yang-Mills theories in four dimensions. In this paper we
generalize previous results obtained by Vafa and Witten \cite{vw} for one of the
twisted theories.

 As in the $\cn=2$ case, the
${\cal R}$-symmetry group of the $\cn=4$ algebra  can be twisted to obtain
a topological model. But since the 
${\cal R}$-symmetry group of the $\cn=4$ theory is $SU(4)$, this topological
twist can be performed in  three  inequivalent ways, so one ends up with three
different TQFTs \cite{vw}\cite{yamron}\cite{ene4}. The twisted theories
are  topological in the sense that the partition  function as well as a
selected set of correlation functions are independent  of the metric which
defines the background geometry. In the short distance  regime,
computations in the twisted theory are given exactly by a saddle-point 
calculation around a certain bosonic background or moduli space, and in
fact  the correlation functions can be reinterpreted  as describing
intersection theory on this moduli space. This correspondence can  be made
more precise through the Mathai-Quillen construction
\cite{ene4}.  Unfortunately, it is not possible to  perform explicit
computations from this viewpoint: the moduli spaces one ends  up with are
generically non-compact, and no precise recipe is known to  properly
compactify them. 

While for the  TQFTs related to asymptotically free ${\cal N}=2$
theories the  interest lies in their ability to define
topological invariants for  four-manifolds, for the twisted
${\cal N}=4$ theories the topological character  is used as a
tool for performing explicit computations which might shed light
on  the structure of the physical ${\cal N}=4$ theory. This
theory is finite and  conformally invariant, and is conjectured
to have a symmetry exchanging strong  and weak coupling and
exchanging electric and magnetic fields, which extends to  a full
$SL(2,\IZ)$ symmetry acting on the microscopic complexified
coupling 
$\tau$ \cite{monoli}. In addition to this, since all the fields
in the theory take values in the  adjoint representation of the
gauge group, it is possible to consider non-trivial gauge
configurations in
$G/Center(G)$ and compute the partition function for  fixed
values of the 't Hooft flux $v\in  H^{2}(X,\pi_1(G))$.
\cite{gthooft} which should behave under $SL(2,\IZ)$
duality in a well-defined  fashion
\cite{gthooft}. This has been checked for the physical $\cn=4$
theory on $T^4$ in \cite{italia}. 
 It is natural
to expect that this property should be  shared by the  twisted
theories on arbitrary four-manifolds. This was checked by Vafa
and  Witten for one of the twisted theories and for gauge group
$SU(2)$ 
\cite{vw}, and it was  clearly mostly interesting to
extend their computation to higher rank groups. Similar results have
been recently derived for another twisted version of the $\cn$=
$4$ theory within the $u$-plane approach \cite{htwist}.

In this paper we will consider the Vafa-Witten theory for gauge 
group $SU(N)$. 
The twisted theory does not contain spinors, so it is  well-defined on
any compact, oriented four-manifold. The ghost-number symmetry  of this
theory is anomaly-free, and therefore the only non-trivial topological 
observable is the partition function itself. As we mentioned above, it
is  possible to consider non-trivial gauge configurations in
$G/Center(G)$ and  compute the partition function for a fixed
value of the 't Hooft flux $v\in  H^{2}(X,\pi_1(G))$. In this
case, however, the Seiberg-Witten approach is not available,
but, as conjectured by Vafa and Witten, one can  nevertheless
compute in terms of the vacuum degrees of freedom of the 
${\cal N}=1$ theory which results from giving bare masses to all
the three  chiral multiplets of the ${\cal N}=4$
theory\footnote{A similar approach was  introduced by Witten in
\cite{wijmp} to obtain the first explicit results for  the
Donaldson-Witten theory just before the far more powerful
Seiberg-Witten approach was available.}. The partition functions 
on $K3$ for gauge group $SU(N)$ and trivial 't Hooft
fluxes have been computed by Vafa and collaborators 
in \cite{estrings}. We will extend their results to 
arbitrary 't Hooft fluxes and compute the partition function on more
general K\"ahler four-manifolds. A brief account of these results has
already appeared in \cite{faro}.

The paper is organized as follows. In sect. 2 we review the structure of
the $\cn=4$ supersymmetric gauge theory in four
dimensions and its topological twisting. In sect. 3 we review the
Vafa-Witten theory, which arises as a twisted version of the $\cn=4$ 
theory, and analyze the vacuum structure of the $\cn=1$ theory which arises by
giving masses to all the three chiral multiplets of the $\cn=4$ theory.
In sect.
4 we derive the partition function on $K3$ for $G=SU(N)$ with prime $N$
and arbitrary values of the 't Hooft fluxes. In sect. 5 we
 generalize the
partition function to more general K\"ahler manifolds and study the 
properties of the resulting formulas under duality and under blow-ups. 
Finally, in sect. 6  we state our conclusions. An appendix deals with a
set of useful identities  and definitions used in the paper.

\newpage
\section{Twisting $\cn=4$ supersymmetric gauge theory on
four-manifolds}
\setcounter{equation}{0}

In this section we  review  some aspects of the four-dimensional
$\cn$=$4$ gauge theory and its topological twisting.

\subsection{The $\cn=4$ supersymmetric gauge theory}

We begin with several well-known remarks concerning the $\cn=4$
supersymmetric gauge theory on flat ${\IR}^4$. 
The $\cn=4$ supersymmetric Yang-Mills
theory is unique once the gauge group
$G$ and the microscopic coupling
$\tau={\frac{\theta}{2\pi}}+{\frac{4\pi^2 i}{e^2}}$ are
fixed. The  microscopic theory contains a gauge or gluon
field,  four chiral spinors (the gluinos) and six real scalars.
All the above fields  are massless and take values in the adjoint
representation of the gauge group $G$. From the point of view of
$\cn=1$ superspace, the theory contains one $\cn=1$ vector 
multiplet and three $\cn=1$ chiral  multiplets. These 
supermultiplets are represented  in
$\cn=1$ superspace  by the superfields
$V$ and $\Phi_s$ ($s=1,2,3$), which  satisfy the  constraints
$V=V^{\dag}$ and $\bar D_\dalpha \Phi_s=0$,  
$\bar D_\dalpha$ being a superspace covariant
derivative\footnote{We follow the  same conventions as in
\cite{ene4}.}.  
%
%
The $\cn=4$ supersymmetry algebra has the automorphism group
$SU(4)_I$,  under which the gauge bosons are scalars, the
gauginos transform in the ${\bf
4}\oplus{\bf\bar 4}$,  and the scalars transform as a self-conjugate 
antisymmetric tensor $\phi_{uv}$ in the ${\bf 6}$. 

The action takes the following form in $\cn=1$ superspace:
\bea
{\cal S}\!\!\! &=&\!\!\! -{\frac{i}{4\pi}}\tau\int d^4 xd^2 \theta\, \tr
(W^2) + {\frac{i}{4\pi}}\bar\tau\int d^4 x d^2
\bar\theta\, \tr (W^{\dag 2}) \ret &+& \!\!\!\! \frac{1}{e^2}
\sum_{s=1}^3 \int d^4 xd^2 \theta d^2
\bar\theta\, \tr(\Phi^{\dag s} \ex^V \Phi_s\ex^{-V}) \ret   
&+&\!\!\!\!\frac{i\raiz}{e^2}
\int d^4x d^2\theta \, \tr\bigl\{\Phi_1[\Phi_2 ,\Phi_3]\bigr\} +
\frac{i\raiz}{e^2}\int d^4 xd^2\bar\theta\,\tr\bigl\{\Phi^{\dag
1}  [\Phi^{\dag 2},\Phi^{\dag 3}]\bigr\}, \ret
\label{cuno}
\eea
where $W_\alpha =-\frac{1}{16}\bar{D}^2 \ex^{-V}D_\alpha
\ex^{V}$ and $\tr$ denotes the trace in the fundamental
representation. 

The theory is invariant under four independent supersymmetries
which transform  under $SU(4)_I$, but only one of these is
manifest in the $\cn=1$ superspace  formulation (\ref{cuno}).
The global symmetry group of $\cn=4$ supersymmetric theories in
${\IR}^4$ is ${\cal H}= SU(2)_L\otimes SU(2)_R\otimes SU(4)_I$,
where  ${\cal K}=  SU(2)_L\otimes SU(2)_R$ is the rotation group
$SO(4)$. The fermionic   generators of the four supersymmetries
are $Q^u{}_\alpha$ and 
$\bar Q_{u\dalpha}$. They transform  as $({\bf 2},{\bf 1},{\bf
\bar 4})
\oplus({\bf 1},{\bf 2},{\bf 4})$ under ${\cal H}$.

The massless $\cn=4$ supersymmetric theory has zero beta
function, and it is  believed  to be exactly finite and
conformally invariant, even non-perturbatively. It is in fact
the most promising candidate for the explicit realization of the
strong-weak coupling duality symmetry conjectured some twenty
years ago by Montonen and Olive \cite{monoli}.

\subsection{Twists of the $\cn=4$ supersymmetric theory}

The twist in the context of four-dimensional
supersymmetric  gauge theories was introduced by Witten in
\cite{tqft}, where it was shown that a twisted version of the
$\cn=2$ supersymmetric gauge theory with gauge  group $SU(2)$ is
a relativistic field-theory representation of the Donaldson 
theory of four-manifolds. In four dimensions, the global symmetry
group of the extended supersymmetric  gauge theories is of the form
$SU(2)_L\otimes SU(2)_R\otimes{\cal I}$,  where ${\cal K}= SU(2)_L\otimes
SU(2)_R$ is the rotation group, and ${\cal I}$ is the chiral
${\cal R}$-symmetry group. The twist can be thought of either as
an  exotic realization of the global symmetry group of the
theory, or as the  gauging (with the spin connection) of a
certain subgroup of the global 
${\cal R}$-current of the theory. 

While in $\cn=2$ supersymmetric gauge theories  the ${\cal
R}$-symmetry group is at most $U(2)$ and thus the twist is
essentially unique (up to an exchange of left and right),  in the
$\cn=4$ supersymmetric gauge theory the  
${\cal R}$-symmetry group is $SU(4)$ and there are three different
possibilities, each  corresponding to a different non-equivalent
embedding of the rotation group into the ${\cal R}$-symmetry
group \cite{vw}\cite{yamron}\cite{ene4}. Two of these possibilities give
rise to topological field theories with two independent BRST-like
topological symmetries. One of these was considered by Vafa and Witten in
\cite{vw}.   The second possibility was first addressed by Marcus
\cite{marcus},  and his analysis was extended in 
\cite{ene4}\cite{blauthomp}. The remaining possibility leads to
the half-twisted theory,  a topological theory with only one BRST
symmetry
\cite{yamron}\cite{ene4}. The generating function of  
topological correlation functions of this theory has been recently
computed for gauge group $SU(2)$ and arbitrary values of the 't
Hooft flux in \cite{htwist} within the $u$-plane framework
\cite{moorewitten}.

\section{The Vafa-Witten theory}

The Vafa-Witten theory can be obtained by twisting the $\cn$=4
supersymmetric gauge theory as described in
\cite{vw}\cite{yamron}\cite{ene4}. The twisted theory has an
anomaly free  Abelian ghost-number symmetry which is a subgroup 
of the $SU(4)_I$ $\C{R}$-symmetry of the $\cn$=$4$ theory.
Therefore,  the partition function is the only non-trivial
topological observable of the theory \cite{vw}. 

The theory has  $2$ independent BRST charges $Q^{\pm}$
of  opposite ghost number. The  field content consists of $3$
scalar fields $\{\phi^{+2},\bar\phi^{-2}, C^{0}\}$, $2$ one-forms
$\{A_{\alpha\dalpha}^{0},
\tilde H_{\alpha\dalpha}^{0}\}$  and $2$ self-dual two-forms
$\{(B^{+}_{\alpha\beta})^0, (H^{+}_{\alpha\beta})^0\}$  on the
bosonic (commuting) side; and 
$2$ scalar fields $\{\zeta^{+1},\eta^{-1}\}$, 
$2$ one-forms $\{\psi_{\alpha\dalpha}^{1},\tilde
\chi_{\alpha\dalpha}^{-1}\}$  and $2$ self-dual two-forms 
$\{(\tilde\psi^{+}_{\alpha\beta})^{+1},
(\chi^{+}_{\alpha\beta})^{-1}\}$   on the fermionic
(anticommuting) side. The superscript stands for the ghost 
number carried by each of the fields. 


The twisted $\cn=4$ supersymmetric action breaks up into a
$Q^{+}$-exact piece  (that is, a  piece which can be written as
$\{Q^{+},{\cal T}\}$, where ${\cal T}$ is a functional  of the fields
of the theory), plus a topological term proportional to  the
instanton number of the gauge configuration,
\begin{equation} {\cal S}_{\mbox{\rm\tiny twisted}}= \{Q^{+},{\cal
T}\}-2\pi ih_v\tau,
\label{ramala}
\end{equation} 
with $h_v$ the instanton number of
a gauge bundle with 't Hooft flux $v$. This is an integer for
$SU(N)$ bundles ($v=0$), but for non-trivial
$SU(N)/\IZ_N$ bundles with $v\not=0$ one has 
\begin{equation}
h_v= -\frac{N-1}{2N}\, v\cdot v\quad\mod~\IZ, 
\label{instnum}
\end{equation}
where $v\cdot v$ stands for $\int_X v\wedge v$ . 
Therefore, as pointed out in 
\cite{vw}, one  would expect the
$SU(N)$ partition function to be invariant under $\tau\to\tau+1$, 
while the $SU(N)/\IZ_N$ theory should be only invariant under
$\tau\to\tau+N$. Notice that, owing to (\ref{ramala}), the partition
function  depends on  the microscopic couplings $e$  and $\theta$
only through the combination $2\pi ih_v \tau$,  and in particular this
dependence is a priori holomorphic (were the orientation of the manifold
$X$ reversed, the partition function would depend  anti-holomorphically
on $\tau$). However there could be
situations in which,  because of some sort of holomorphic anomaly,
the partition function would  acquire an explicit anomalous
dependence on $\bar\tau$. This seems to  be the case, for
example, for the theory defined on 
$\cp^{2}$ \cite{vw} and, more generally, on manifolds with
$b^{+}_2=1$ \cite{estrings}. Somewhat related results have been derived
for the Donaldson-Witten theory in the context of the $u$-plane 
formalism \cite{moorewitten}.


\subsection{Mass perturbations and reduction to $\cn$=$1$}
It is a well-known fact that on complex manifolds the exterior
differential $d$ splits into the Dolbeaut operators $\partial$ 
and $\bar\partial$. 
In a similar way, as pointed out in \cite{wijmp}, on a  K\"ahler
manifold the number of BRST charges of a twisted supersymmetric
theory is doubled, in such a way that, for example,  the 
Donaldson-Witten theory has an enhanced
$\cn_T=2$ topological symmetry   on K\"ahler manifolds, while the
Vafa-Witten theory has $\cn_T=4$ topological  symmetry. In each
case, one of the BRST charges comes from the underlying $\cn=1$ 
subalgebra which corresponds to the formulation of the physical
theory in 
$\cn=1$ superspace. By suitably adding mass terms for some of the 
chiral superfields in the theory, one can break the extended
($\cn=2$ or
$\cn=4$)  supersymmetry of the physical theory down to $\cn=1$.
For the reason sketched  above, the corresponding twisted massive
theory on K\"ahler manifolds should still retain at least  one
topological symmetry. One now exploits the metric independence of
the  topological theory. By scaling up the metric in the
topological theory, 
$g_{\mu\nu}\to tg_{\mu\nu}$, one can take the limit $t\to\infty$.
In this  limit, the metric on $X$ becomes nearly flat, and it is 
reasonable that the computations in the topological field  theory
can be performed in terms of the  vacuum structure of the $\cn=1$
theory. 

One could wonder as to what the effect of the perturbation may
be. The introduction of a mass perturbation may  (and in general
will) distort the original topological field theory.  This poses
no problem in the case of the  Donaldson-Witten theory, as Witten
was able to prove that the perturbation is topologically trivial,
in the sense that it affects the theory in an important but
controllable way \cite{wijmp}.  As for the Vafa-Witten theory
\cite{vw}\cite{masas}\cite{coreatres},  the twisted massive
theory is topological on K\"ahler four-manifolds  with
$h^{2,0}\not=0$, and the partition function is actually invariant
under the perturbation. The constraint $h^{2,0}\not=0$ comes
about as follows.  In the twisted theory  the chiral superfields
of the $\cn=4$ theory are no longer scalars, so the  mass terms
can not be invariant under the holonomy group of the manifold 
unless one of the mass parameters be a holomorphic two-form
$\omega$. 

The massive $\cn=1$ theory has the tree level superpotential 
\begin{align}
\C{W}&= \frac{i\raiz}{e^2}
\int d^4x d^2\theta \, \tr\bigl\{\Phi_1[\Phi_2 ,\Phi_3]\bigr\}
+m\int_X d^2\theta\tr{(\Phi_1\Phi_2)}
\ret
&+\int_X d^2\theta\omega\tr{(\Phi_3)^2}+ \text{h.c.}
\ret
\label{perita}
\end{align}
Up to a constant rescaling the equations for a critical 
point of $\C{W}$ are
\begin{align}
[\Phi_3,\Phi_1]&=-m\Phi_3,\ret
[\Phi_3,\Phi_2]&=m\Phi_2,\ret
[\Phi_1,\Phi_2]&=2\omega\Phi_3.\ret &{}
\la{critical}
\end{align}

As noted in \cite{vw}\cite{donagi}, these equations are the
commutation relations of the Lie algebra of $SU(2)$, and
therefore the classical vacua of the resulting $\cn=1$ theory can
be classified by the complex conjugacy classes of homomorphisms
of the
$SU(2)$ Lie algebra to that of $G=SU(N)$.

 Eqs. \eqs{critical}   admit a trivial solution
$\Phi_1\,$=$\,\Phi_2\,$=$\,\Phi_3\,$=$\,0$ where  the gauge group
is unbroken and which reduces at low energy to  the $\cn\,$=$\,1$
pure
$SU(N)$ gauge theory (which has $N$ discrete vacua), and a non-trivial
solution (the irreducible embedding in \cite{vw}) where the gauge
group is completely broken. This corresponds to
$\{\Phi_1,\Phi_2,\Phi_3\}$ defining the  representation $N$ of
$SU(2)$. All these vacua have a mass gap: the irreducible
embedding is a Higgs vacuum, while the presence of a mass gap in
the trivial vacua is a well-known feature of the $\cn=1$ theory.
When $N$ is prime, these are the only relevant vacua of the
$\cn=1$ theory. There are other, more general, solutions to 
\eqs{critical} which leave different subgroups of $G$ unbroken.
However, in all these solutions the unbroken gauge group contains
$U(1)$ factors and one expects on general grounds that they
should not contribute to the partition function \cite{vw}. 
On the other hand, when $N$ is not prime, there are additional
contributions coming from embeddings for which the unbroken gauge 
group is $SU(d)$, where $d$ is a positive divisor of $d$. The
low-energy theory is again an $\cn=1$ $SU(d)$ gauge theory wihout
matter with $d$ massive discrete vacua. 

In the long-distance limit, the partition function
is given as a finite sum over the contributions of the discrete
massive vacua of  the resulting
$\cn=1$ theory.  For
$G=SU(N)$  the number of such vacua is given by the sum of the
positive divisors of $N$ 
\cite{donagi}. The contribution of each vacuum is universal
(because of the  mass gap), and can be fixed by comparing to
known mathematical results \cite{vw}.

\newpage

\section{The partition function on $K3$}

As a first step towards the derivation of the formula for the
partition  function we will consider the theory on $K3$, where
some explicit results are already available. For $X$ a $K3$
surface the canonical divisor is trivial, so there exists a
nowhere vanishing section of the bundle of $(2,0)$ forms.
Therefore, the mass perturbation $\omega$   does not vanish
anywhere and the above analysis of the vacuum  structure of the
$\cn$=$1$ theory carries over without change. 

 The structure of the partition function for trivial 't Hooft
flux was conjectured in \cite{vw}. This conjecture has been
confirmed in
\cite{estrings} by studying the effective theory on $N$
coincident $M5$-branes wrapping around $K3\times T^2$. The
partition function for zero 't Hooft flux is {\sl almost} a Hecke
transformation  of order $N$ \cite{apostol} of
$G(\tau)=\eta(\tau)^{-24}$, with
$\eta(\tau)$ the Dedekind function  -- see eq. (3.7) in
\cite{estrings}:    

\begin{equation} 
Z_{v=0}\equiv Z_N=\frac{1}{N^2}\sum_{ \genfrac{}{}{0pt}{2}{
0\leq a,b,d\in\IZ}{ad=N,~  b<d}}
d\, G\left(\frac{a\tau+b}{d}\right).
\label{suocho}
\end{equation}
Notice that the number of terms in \eqs{suocho} equals the sum of
the positive divisors of $N$ as we mentioned above. When $N$ is
prime the formula is considerably simpler 
\begin{equation}  Z_{v=0}=\frac{1}{N^2}G(N\tau) + \frac{1}{N}
\sum_{m=0}^{N-1}\,
 G\left(\frac{\tau+m}{N}\right).
\label{suochoo}
\end{equation} There are $N+1$ terms, the first one corresponding
to the irreducible embedding, and the other $N$ to the vacua of
the $\cn$=$1$
$SU(N)$ SYM theory.

The $SU(N)$ partition function is defined from \eqs{suocho} as
$Z_{SU(N)}=
\frac{1}{N} Z_{v=0}$. From it, the $SU(N)/\IZ_N$ partition
function
$Z_{SU(N)/\IZ_N}= \sum_v Z_v$ can be obtained via a modular
transformation \cite{vw} (see the appendix for details)  

%
%

\begin{equation}
 Z_{SU(N)/\IZ_N}(\tau)= N^{\chi/2}\left(\frac{\tau}{i}
\right)^{\chi/2}Z_{SU(N)}(-1/\tau)=
\frac{1}{N^2}\sum_{ \genfrac{}{}{0pt}{2}{a,b,d}{p=\gcd\,(b,d)}} 
d^{12} p^{11}
G\left(\frac{a\tau+b}{d}\right).
\label{sudiez}
\end{equation}
Notice the first equality in \eqs{sudiez},  
which is, up to some correction factors which vanish in flat
space,  the original Montonen-Olive conjecture. 

To generalize \eqs{suochoo} for gauge configurations with
arbitrary 't Hooft flux we proceed as in \cite{vw}. The $N$
contributions coming from the $N=1$ pure gauge theory vacua are
related by an anomalous  chiral symmetry which takes
$\tau\to\tau+1$. The anomaly is 
$2Nh_v-(N^2-1)\left(\frac{\chi+\sigma}{4}\right)=-(N-1)v\cdot
v+\cdots$,  which is half the anomaly in Donaldson-Witten theory.
Hence,  the contributions from each vacuum pick anomalous phases 
$\ex^{-i\pi m h_v}=\ex^{i\pi\frac{N-1}{N}m v^2}$.  As for the
contribution coming from the irreducible embedding, modular
invariance  requires that  it vanishes unless $v=0$. Hence, 

\begin{equation} Z_v = \frac{1}{N^2} G(N\tau)\delta_{v,0} +
\frac{1}{N}
\sum_{m=0}^{N-1}\,
\ex^{i\pi\frac{N-1}{N}m v^2}\, G\left(\frac{\tau+m}{N}\right). 
\la{sudiezsiete}
\end{equation} The $Z_v$ transform into each other under the
modular group as predicted in 
\cite{vw} 
\bea Z_v(\tau+1)\!\!\! &=&\!\!\!\ex^{-i\pi\frac{N-1}{N}v^2} 
Z_v(\tau), 
\ret\ret  Z_v(-1/\tau)\!\!\! &=&\!\!\! N^{-11}\left(\frac{\tau}{i}
\right)^{-24}
\sum_u \ex^{\frac{2i\pi u\cdot v}{N}} Z_u(\tau). \ret 
\label{sucuatro2}
\eea To evaluate the sum over $u$ we use formulas \eqs{terran3}
and
\eqs{terran4} in the appendix\footnote{Note that $K3$ has
$\chi=24$,
$\sigma=-16$, $b_1=0$ and $b_2=22$.}. 

%
%

By summing over $v$ in \eqs{sudiezsiete} we can check
\eqs{sudiez} 
\bea Z_{SU(N)/\IZ_N}\!\!\!&=&\!\!\! \sum_v Z_v
\ret\!\!\!&=&\!\!\!\frac{1}{N^2} G(N\tau)+ N^{21} G(\tau/N) +
N^{10}\sum_{m=1}^{N-1}\,
 G\left(\frac{\tau+m}{N}\right). \ret
\la{supino}
\eea
The above results only hold for prime $N$. The appropriate generalization
for arbitrary $N$ should be also investigated.

\hyphenation{Mont-o-nen}

\newpage
\section{More general K\"ahler manifolds}

On more general K\"ahler manifolds the spatially dependent mass
term vanishes where $\omega$ does, and  we will assume as in
\cite{vw}\cite{wijmp} that $\omega$  vanishes with multiplicity
one on a union of disjoint,  smooth complex curves $C_j$,
$j=1,\ldots n$ of genus $g_j$ which represent the  canonical
divisor $K$ of $X$. The vanishing of $\omega$ introduces
corrections  involving $K$ and additional modular functions 
whose precise form is not known a priori. In the $G=SU(2)$ case, 
each of the
$\cn=1$ vacua bifurcates along each of the components $C_j$ of
the  canonical divisor into two strongly coupled massive vacua.
This vacuum  degeneracy is believed to stem \cite{vw}\cite{wijmp}
from the spontaneous  breaking of a $\IZ_2$ chiral symmetry which
is unbroken in bulk. This is exactly the same pattern that arises
in all known examples of twisted $\cn=2$ theories with gauge
group $SU(2)$ as the Donaldson-Witten theory and its 
generalizations \cite{wijmp}\cite{moorewitten}\cite{polynom}. 
This in turn
seems to be related to the possibility of rewritting the corrections
near the canonical divisor in terms of the Seiberg-Witten
invariants
\cite{monopole}. In fact, it is known that the Vafa-Witten
partition function for $G=SU(2)$ can be rewritten in terms of the
Seiberg-Witten  invariants \cite{coreatres}.

The form of the corrections for $G=SU(N)$ is more involved. From
related results on Donaldson-Witten theory \cite{mmtwo} we know
that  the higher-rank case presents some new features. We have
not been able to disentangle the structure of the vacua near the
canonical  divisor from first principles. Instead, we will
exploit the expected behaviour of the partition function under
blow-ups of $X$. This, together with the modular invariance of
the partition function will suffice to completely determine the
unknown functions.

\subsection{Behaviour under blow-ups}

Blowing up a point on a K\"ahler manifold $X$  replaces it  with
a new K\"ahler manifold
$\hat X$ whose second cohomology lattice is 
$H^{2}({\hat X},\IZ)= H^{2}({X},\IZ)\oplus I^{-}$, where $I^{-}$
is the  one-dimensional lattice spanned by the Poincar\'e dual of
the exceptional divisor 
$B$ created by the blow-up. Any allowed $\IZ_N$ flux $\hat v$ on
$\hat X$  is of the form $\hat v=v\oplus r$, where $v$ is a  flux
in $X$ and $r=\lambda B$, $\lambda=0,1,\ldots N-1$. The main
result concerning the $SU(2)$ partition function in \cite{vw} is that
under blowing up a point on a K\"ahler  four-manifold with
canonical divisor as above, the partition functions for  fixed 't
Hooft fluxes $\hat Z_{\hat X,\hat v}$ factorize as $Z_{X,v}$
times  a level $1$ character of the $SU(2)$ WZW model. It would be
natural to expect that the same factorization holds for $G=SU(N)$, 
but now with the level 1 $SU(N)$ characters. 
In fact, the same behaviour under blow-ups  has been proved by  
Yoshioka \cite{yoshioka} for the generating function of Euler
characteristics of $SU(N)$ instanton moduli space on K\"ahler
manifolds. This should not come out as a surprise since it is
known  that, on certain four-manifolds, the partition function
of  Vafa-Witten theory computes Euler characteristics of 
instanton moduli spaces \cite{vw}\cite{estrings}. This can be
confirmed  by realizing the Vafa-Witten theory as the low-energy
theory of
$M5$-branes wrapped on $X\times T^2$ \cite{dijkfvbr}. It seems   
therefore natural to assume that the same factorization holds for
the partition function with $G=SU(N)$. Explicitly, given a 't
Hooft flux 
$\hat v=v\oplus \lambda B$, $\lambda=0,1,\ldots N-1$, on $\hat
X$,  we assume the factorization \cite{yoshioka}
\begin{equation} 
Z_{\hat X,\hat v}(\tau)=
Z_{X,v}(\tau)\,\,\frac{\chi_{\lambda}(\tau)}{\eta(\tau)},
\label{factoriz}
\end{equation} 
where $\chi_{\lambda}(\tau)$ is the appropriate level $1$
character of
$SU(N)$ -- see Appendix A.3 for details. This assumption fixes
almost completely the form of the partition functions. Some loose
ends can be tied up by demanding modular invariance of the
resulting expression.

\subsection{The formula for the partition function}

Given the assumptions above, and taking into account the
structure of the partition function on $K3$, we are in a position
to write down the  formula for  K\"ahler four-folds $X$ with
$h^{(2,0)}\not=0$. We will first assume that the canonical
divisor $K$ is connected and with genus 
$g-1=2\chi+3\sigma$. The formula is then

\bea
Z_v =&&\!\!\!\!\!\!\!\!\! \left (\sum_{\lambda=0}^{N-1}
\left(\frac{\chi_{\lambda}}{\eta}
\right)^{1-g} \delta_{v,\lambda [K]_N}\right)\left 
(\frac{1}{N^2} G(N\tau)\right)^{\nu/2} 
\ret &&\!\!\!\!\!\!\!\!\!\!\!\!\!\!+ N^{1-b_1}
\sum_{m=0}^{N-1}\left( \sum_{\lambda=0}^{N-1} 
\left(\frac{\chi_{m,\lambda}}{\eta}\right)^{1-g} 
\ex^{\frac{2i\pi}{N}\lambda v\cdot[K]_N}\right)
\ex^{i\pi\frac{N-1}{N}m v^2}\left( \frac{1}{N^2}
G\left(\frac{\tau+m}{N}\right)\right)^{\nu/2}, \ret
\label{suno}
\eea
where $\nu$=$\frac{\chi+\sigma}{4}$, $G(\tau)$=$\eta(\tau)^{-24}$
(with $\eta$  the Dedekind function) and $[K]_N$
is the reduction modulo $N$ of the Poincar\'e dual of $K$. In (\ref{suno})
$\chi_\lambda$ are the
$SU(N)$ characters at level $1$ (see Appendix A.3)  and
$\chi_{m,\lambda}$  are certain linear combinations thereof
\begin{equation}
\chi_{m,\lambda}(\tau)=\frac{1}{N} \sum_{\lambda'=0}^{N-1}
\ex^{-\frac{2 i\pi}{N}
\lambda\lambda'}\ex^{i\pi\frac{N-1}{N}m (\lambda')^2}
\chi_{\lambda'}(\tau),\qquad0\leq
m,\lambda\leq N-1.
\label{suonce2}
\end{equation}

The structure of the corrections near the canonical divisor in 
\eqs{suno} suggests that the mechanism at work in this case is
not chiral symmetry  breaking. Indeed, near $K$ there is an
$N$-fold bifurcation of the vacuum, and the functions
$\chi_\lambda$, $\chi_{m,\lambda}$ (with $m$ fixed) are not
related  by a shift in $\tau$ as it would be the case were chiral
symmetry breaking responsible for the bifurcation.  A plausible
explanation for this bifurcation could be found in the spontaneous
breaking of the center of the gauge group (which for $G=SU(N)$
is  precisely $\IZ_N$.)  This could come about as follows. Let
us focus on the irreducible embedding.  For trivial canonical
divisor the gauge group is almost but not completely Higgsed in
this vacuum. In fact, since the scalar fields transform in the
adjoint representation of $SU(N)$, the center $\IZ_N\subset
SU(N)$ remains unbroken. The $SU(N)$ gauge threory has 
$\IZ_N$ string-like solitons \cite{gthooft} which carry
non-trivial $\IZ_N$-valued electric and magnetic quantum numbers.
If these solitons condense, the center $\IZ_N$ is completely
broken giving rise to an $N$-fold degeneracy of the vacuum. Each
vacuum is singled out by a different value of the $\IZ_N$-valued
flux. Now for non-trivial canonical divisor
$K$ as above, the irreducible vacuum separates into
$N$ vacua with magnetic fluxes $\lambda[K]_N$! One could be
tempted to speculate further and identify the surface $K$
(or the $C_j$ below) with the world-sheet of the condensed
string soliton.       

As in \cite{vw} we can generalize the above formula for the case
that  the canonical divisor consists of $n$ disjoint smooth
components 
$C_j$, $j=1,\ldots,n$ of genus $g_j$ on which $\omega$ vanishes
with multiplicity one. The resulting expression is:

\begin{align}
&Z_v = \left (\sum_{\vec\varepsilon}
\delta_{v,w_N(\vec\varepsilon\,)}
\prod_{j=1}^{n}\prod_{\lambda=0}^{N-1} 
\left(\frac{\chi_{\lambda}}{\eta}
\right)^{(1-g_j)\delta_{\varepsilon_j,\lambda}}\right)
\left (\frac{1}{N^2} G(N\tau)\right)^{\nu/2} 
\ret &+ N^{1-b_1}
\sum_{m=0}^{N-1}\left[\prod_{j=1}^{n}
\left( \sum_{\lambda=0}^{N-1} 
\left(\frac{\chi_{m,\lambda}}{\eta}\right)^{1-g_j} 
\ex^{\frac{2i\pi}{N}\lambda v\cdot[C_j]_N}\right)\right]
\ex^{i\pi\frac{N-1}{N}m v^2}\left( \frac{1}{N^2}
G(\frac{\tau+m}{N})\right)^{\nu/2}, 
\ret 
\label{sudos}
\end{align}
where $q={\hbox{\rm exp}}(2\pi i\tau)$, $\alpha={\hbox{\rm
exp}}(2\pi i/N)$,
$[C_j]_N$ is the reduction modulo $N$ of the Poincar\'e dual of
$C_j$,  and 

\begin{equation} 
w_N(\vec\varepsilon)=\sum_{j} \varepsilon_j [C_j]_N,
\label{sutres}
\end{equation}

where $\varepsilon_j=0,1,\ldots N-1$ are chosen independently. Notice that
\eqs{sudos} reduces to \eqs{suno} when $n=1$.

The formulas for the partition function do not apply directly to the $N=2$
case. For 
$N=2$ there are some extra relative phases $t_i$ -- see equations 
(5.45) and
(5.46)  in \cite{vw} -- which are absent for $N>2$ and prime. 
Modulo these extra  phases, (\ref{suno}) and \eqs{sudos} are a
direct generalization of Vafa and Witten's  results. They reduce
on $K3$ to  the formula of Minahan,  Nemeschansky, Vafa and
Warner \cite{estrings}  and generalize their results to non-zero
't Hooft flux. 

\subsubsection{Blow-ups} 

Given \eqs{sudos}, we can see explicitly how the factorization 
property \eqs{factoriz} works. Let $X$ be a K\"ahler four-fold with 
Euler characteristic $\chi=2(1-b_1)+b_2$, signature $\sigma=b_2^{+}-
b_2^{-}$ and canonical divisor $K=\cup_{j=1}^{n}C_j$, and let $\hat X$ be
its  one blow-up at a smooth point. Then $\hat b_1=b_1$, $\hat
b_2=b_2+1$,   
$\hat\chi=\chi+1$, $\hat\sigma=\sigma-1$ and $\hat K=K\cup B$, where $B$ is 
the exceptional divisor, which satisfies $B\cdot C_j=0$ and 
$B^2=-1=g_B-1$. Consider a 't Hooft  flux $\hat v=v\oplus \hat\lambda B$
in $\hat X$, where $v$ is a flux in $X$ and 
$\hat\lambda$ is an integer defined modulo $N$. Now $\hat \nu=\nu$, $\hat v^2=
v^2-{\hat\lambda}^2$, 
$\hat v\cdot C_j=v\cdot C_j$, $\hat v\cdot B=\hat\lambda B^2=-\hat\lambda$ and 
$\hat w_N(\vec\epsilon)=
\sum_{j=1}^{n} \varepsilon_j [C_j]_N+\hat\varepsilon \,B$. 
Thus, the partition function 
\eqs{sudos} takes the form 
\begin{align}
\hat Z_{\hat X,\hat v} &= \left (\sum_{\vec\varepsilon,\hat\varepsilon}
\delta_{v,w_N(\vec\varepsilon\,)}\delta_{\hat\lambda,\hat\varepsilon}
\prod_{j=1}^{n}\prod_{\lambda=0}^{N-1} 
\left(\frac{\chi_{\lambda}}{\eta}
\right)^{(1-g_j)\delta_{\varepsilon_j,\lambda}}
\left(\frac{\chi_{\lambda}}{\eta}
\right)^{(1-g_B)\delta_{\hat\varepsilon,\lambda}}\right)
\left (\frac{1}{N^2} G(q^N)\right)^{\nu/2} 
\ret &+ N^{1-b_1}
\sum_{m=0}^{N-1}\left[\prod_{j=1}^{n}
\left( \sum_{\lambda=0}^{N-1} 
\left(\frac{\chi_{m,\lambda}}{\eta}\right)^{1-g_j} 
\ex^{\frac{2i\pi}{N}\lambda v\cdot[C_j]_N}\right)
\left( \sum_{\lambda=0}^{N-1} 
\left(\frac{\chi_{m,\lambda}}{\eta}\right)^{1-g_B} 
\ex^{-\frac{2i\pi}{N}\lambda\hat\lambda}\right)\right]\ret 
&\ex^{i\pi\frac{N-1}{N}m v^2}\ex^{-i\pi\frac{N-1}{N}m {\hat\lambda}^2}
\left( \frac{1}{N^2} G(\alpha^m
q^{1/N})\right)^{\nu/2}, \ret
\la{suudos}
\end{align}
and therefore
\begin{align}
\hat Z_{\hat X,\hat v} &=\left(\frac{\chi_{\hat\lambda}}{\eta}
\right) \left (\sum_{\vec\varepsilon}
\delta_{v,w_N(\vec\varepsilon\,)}
\prod_{j=1}^{n}\prod_{\lambda=0}^{N-1} 
\left(\frac{\chi_{\lambda}}{\eta}
\right)^{(1-g_j)\delta_{\varepsilon_j,\lambda}}\right)
\left (\frac{1}{N^2} G(q^N)\right)^{\nu/2} 
\ret &+ N^{1-b_1}
\sum_{m=0}^{N-1}
\left( \sum_{\lambda=0}^{N-1} 
\left(\frac{\chi_{m,\lambda}}{\eta}\right) 
\ex^{-\frac{2i\pi}{N}\lambda\hat\lambda}\ex^{-i\pi\frac{N-1}{N}m
{\hat\lambda}^2}\right)
\left[\prod_{j=1}^{n}
\left( \sum_{\lambda=0}^{N-1} 
\left(\frac{\chi_{m,\lambda}}{\eta}\right)^{1-g_j} 
\ex^{\frac{2i\pi}{N}\lambda v\cdot[C_j]_N}\right)
\right]\ret 
&\ex^{i\pi\frac{N-1}{N}m v^2}
\left( \frac{1}{N^2} G(\alpha^m
q^{1/N})\right)^{\nu/2}.\ret 
\label{sudos2}
\end{align}
Now, from \eqs{suonce2} it follows that 
\begin{equation}
 \sum_{\lambda=0}^{N-1} 
\left(\frac{\chi_{m,\lambda}}{\eta}\right) 
\ex^{-\frac{2i\pi}{N}\lambda\hat\lambda}\ex^{-i\pi\frac{N-1}{N}m
{\hat\lambda}^2}=\frac{1}{N}\sum_{\lambda,\lambda'} 
\ex^{-\frac{2 i\pi}{N}
\lambda(\lambda'+\hat\lambda)}
\ex^{i\pi\frac{N-1}{N}m ((\lambda')^2-{\hat\lambda}^2)}
\left(\frac{\chi_{\lambda'}}{\eta}\right).
\la{suonce3}
\end{equation}
Summing over $\lambda$ and using \eqs{prono} we get 
\begin{align}
\frac{1}{N}&\sum_{\lambda,\lambda'} 
\ex^{-\frac{2 i\pi}{N}
\lambda(\lambda'+\hat\lambda)}
\ex^{i\pi\frac{N-1}{N}m ((\lambda')^2-{\hat\lambda}^2)}
\left(\frac{\chi_{\lambda'}}{\eta}\right)\ret&=
\sum_{\lambda'}\delta_{\lambda'+\hat\lambda,0}\, 
\ex^{i\pi\frac{N-1}{N}m ((\lambda')^2-{\hat\lambda}^2)}
\left(\frac{\chi_{\lambda'}}{\eta}\right)\ret
&=\frac{\chi_{-\hat\lambda}}{\eta}=\frac{\chi_{N-\hat\lambda}}{\eta}=
\frac{\chi_{\hat\lambda}}{\eta}.
\la{suonce4}
\end{align}
Hence, 
\begin{align}
\hat Z_{\hat X,\hat v} &=\left(\frac{\chi_{\hat\lambda}}{\eta}
\right) \left (\sum_{\vec\varepsilon}
\delta_{v,w_N(\vec\varepsilon\,)}
\prod_{j=1}^{n}\prod_{\lambda=0}^{N-1} 
\left(\frac{\chi_{\lambda}}{\eta}
\right)^{(1-g_j)\delta_{\varepsilon_j,\lambda}}\right)
\left (\frac{1}{N^2} G(q^N)\right)^{\nu/2} 
\ret &+ N^{1-b_1}
\sum_{m=0}^{N-1}
\left(\frac{\chi_{\hat\lambda}}{\eta}
\right)
\left[\prod_{j=1}^{n}
\left( \sum_{\lambda=0}^{N-1} 
\left(\frac{\chi_{m,\lambda}}{\eta}\right)^{1-g_j} 
\ex^{\frac{2i\pi}{N}\lambda v\cdot[C_j]_N}\right)
\right]\ret 
&\ex^{i\pi\frac{N-1}{N}m v^2}
\left( \frac{1}{N^2} G(\alpha^m
q^{1/N})\right)^{\nu/2}=\left(\frac{\chi_{\hat\lambda}}{\eta}
\right) Z_{X,v},\ret 
\label{sudos3}
\end{align}
as expected. 

\subsubsection{Modular transformations}
We will now study the modular properties of the partition functions 
\eqs{suno} and \eqs{sudos}.  With the formulas in the
appendix one can  check that they have the expected  modular 
behaviour\footnote{We assume as in \cite{vw} that there is no
torsion in $H_2(X,\iz)$. Were this not case, Eqs. \eqs{sucinco}
and \eqs{sudos} above should be modified along the lines 
explained in \cite{wiads2}.}

\begin{align}
Z_v(\tau+1)
&=\ex^{\frac{i\pi}{12}N(2\chi+3\sigma)}
\ex^{-i\pi\frac{N-1}{N}v^2} 
Z_v(\tau), 
\ret 
Z_v(-1/\tau) &= N^{-b_2/2}\left(\frac{\tau}{i}
\right)^{-\chi/2}
\sum_u \ex^{\frac{2i\pi u\cdot v}{N}} Z_u(\tau), 
\la{sucuatro}\\ 
\intertext{and also, with  $Z_{SU(N)}= N^{b_1-1} Z_{0}$ and 
$Z_{SU(N)/\IZ_N}=\sum\limits_{v} Z_v$,} 
Z_{SU(N)}(\tau+1)&=\ex^{\frac{i\pi}{12}N(2\chi+3\sigma)}
Z_{SU(N)}(\tau),\ret
Z_{SU(N)/\IZ_N}(\tau+N)&=\ex^{\frac{i\pi}{12}N^2(2\chi+3\sigma)}
Z_{SU(N)/\IZ_N}(\tau), \la{sucinco1}\\
\intertext{and}
Z_{SU(N)}(-1/\tau)&= N^{-\chi/2}\left(\frac{\tau}{i}
\right)^{-\chi/2}Z_{SU(N)/\IZ_N}(\tau),
\label{sucinco}
\end{align}
which is the Montonen-Olive relation. Notice that since $N$ is
odd,  the
$SU(N)$ (or $SU(N)/\IZ_N$) partition function is modular (up to a
phase) for $\Gamma_0(N)$, or $\Gamma_0(N/2)$ for spin manifolds.
On the other hand, for {\sl even} $N$ one would expect on 
general grounds \cite{vw} modularity for 
$\Gamma_0(2N)$, or at most $\Gamma_0(N)$ for spin manifolds.

\subsubsection{The partition function on $T^4$}
We will finish by considering the twisted theory on  
$T^4$, where an unexpected result emerges. As $K3$, $T^4$ is a 
compact hyper-K\"ahler manifold (hence with trivial canonical 
divisor). It has $b_1\,$=$\,4$, $b_2\,$=$\,6$ and $\chi\,$=
$0\,$=$\,\sigma$.  
On $T^4$ the partition function \eqs{suno} reduces to its bare
bones
\begin{equation}
Z_v = \delta_{v,0} + \frac{1}{N^3}
\sum_{m=0}^{N-1}\,
\ex^{i\pi\frac{N-1}{N}m v^2},
\la{sudiezsiete2}
\end{equation}
and does not depend on $\tau$! This should be compared with the
formulas in \cite{italia}. The $Z_v$ are self-dual in the
following sense 
\begin{equation}
Z_v =\frac{1}{N^3}
\sum_u \ex^{\frac{2i\pi u\cdot v}{N}} Z_u. 
\la{sumatra}
\end{equation}

\newpage
\section{Conclusions}

In this paper we have obtained the partition function of the Vafa-Witten theory 
for gauge group $SU(N)$ (with prime $N$) on K\"ahler four-manifolds with
$b_2^{+}>1$. The resulting formulas (\ref{suno}) and
(\ref{sudos}) turn out to transform as expected under the modular group, and they 
 can be seen as predictions for the Euler numbers of instanton
moduli spaces on those four-manifolds. 

It could be interesting to investigate
whether
\eqs{suno} and
\eqs{sudos} can be rewritten in terms of the Seiberg-Witten invariants. 
We believe that this is not the case for the following reason. 
Let us suppose that it is actually possible to do so. Then one would
expect, by analogy with the result for $SU(2)$ \cite{coreatres},that the 
Donaldson-Witten partition function for $SU(N)$ \cite{mmtwo} should be
recovered from the Vafa-Witten $SU(N)$ partition function in the
decoupling limit $q\to 0$, $m\to\infty$ with $m^4q$ fixed.  In particular,
one would expect that the structure of the corrections involving the
canonical divisor should be preserved in this limit. Now in the DW
partition function in
\cite{mmtwo}, these corrections are written
in terms of the Seiberg-Witten classes $x$ 
\cite{monopole}. 
For $G=SU(N)$ these basic classes appear in the generic form
$\sum_{x_1,\ldots  x_{N-1}}n_{x_1}\cdots n_{x_{N-1}}$ ($n_{x_l}$ are the
Seiberg-Witten invariants \cite{monopole}). Therefore, for $G=SU(N)$ there
are $N-1$ independent basic classes contributing to the above sum. On a
K\"ahler manifold with canonical divisor $K=C_1\cup C_2\cup\cdots\cup
C_n$, with the $C_j$ disjoint and with multiplicity one, each of these
basic classes can be written as
\begin{equation}
x_l=\sum\nolimits_{\rho_l^j}
\rho_l^j C_j,
\notag
\end{equation}
with each $\rho_l^j=\pm 1$ \cite{monopole}, and the sum over
the basic classes can be traded for a sum over the $\rho_l^j$. This is
analogous to the sum over the $\varepsilon_j$ in \eqs{sudos}, and both
sums should contain the same number of terms were it possible to rewrite
\eqs{sudos} in terms of the basic classes. However, while in the sum over
the
$\rho_l^j$ there are 
$2^{n(N-1)}$ terms, the sum over the $\varepsilon_j$ contains $N^n$ terms. 
Notice that these two numbers do coincide when $N=2$, as it should be, but
for
$N\not=2$ this is no longer the case. 

It would certainly be mostly interesting to extend these results to all 
$N$ (not necessarily prime), and to investigate what the large $N$ limit 
of \eqs{suno} and \eqs{sudos} correspond to on the gravity side in the
light of the AdS/CFT correspondence \cite{malda}. Although there are already some
indications of how this correspondence should work \cite{gopaku}\cite{hull}, a
clear understanding is still lacking. We expect to address some of these issues in
the near future.

\vskip2cm
\begin{center} {\bf Acknowledgments}
\end{center}

\vspace{4 mm}

We would like to thank M. Mari\~no, A.V. Ramallo and K. Yoshioka
for helpful discussions. C.L. would like to thank the
organizers of the ``Lisbon School on Superstrings," where some
of these results were announced. This work was supported in part
by  DGICYT under grant PB96-0960, and by the EU Commission under
TMR grant FMAX-CT96-0012.

\vfil
\newpage

{\noindent\Large\bf Appendix}
\def\theequation{A.\arabic{equation}}
\def\thesubsection{A.\arabic{subsection}}
\setcounter{equation}{0}
\vskip1.5cm

Here we collect some useful formulas which should help the 
reader follow the computations in the paper.
\subsection{Modular forms}

The function $G$ is defined as 
\begin{equation}
G(\tau)=\frac{1}{\eta(\tau)^{24}},
\la{lage}
\end{equation}
and is a modular form of weight $-12$
\begin{equation}
G(\tau)\,\mapright{\tau\to\tau+1}\, 
G(\tau),\qquad\qquad G(\tau)\,\mapright{\tau\to
-1/\tau}   \tau^{-12}  G(\tau),
\la{supadre}
\end{equation}
From \eqs{supadre} we can determine the modular behaviour of the
different modular forms in the $K3$ partition function 

\begin{align} 
G(N\tau)&\mapright{\tau\to -1/\tau}\tau^{-12} N^{12}
G(\tau/N),\ret \ret
G\left(\frac{\tau+m}{N}\right)&\mapright{\tau\to
-1/\tau}
\tau^{-12}G\left(\frac{\tau+h}{N}\right),\ret &{}
\label{susi}
\end{align}
where $1\leq h \leq N-1$, $mh=-1~\mod~ N$ and $N$ prime. 

For arbitrary $N$ one has to consider the modular forms
$G\left(\frac{a\tau+b}{d}\right)$, where $ad =N$ and $b< d$
\cite{estrings}. These functions transform as follows
\begin{equation}
G\left(\frac{a\tau+b}{d}\right)\mapright{\tau\to
-1/\tau}\tau^{-12}
\left(\frac{a}{p}\right) ^{-12}G\left(\frac{p\tau+ab'}{a\tilde
d}\right),
\label{susiete}
\end{equation}
where $p=\gcd~ (b,d)$, $\tilde d=d/p$, $\tilde b=b/p$, $b'\tilde
b=-1~
\mod~ \tilde d$. If $b=0$, then $p=d$ and $b'=0=\tilde b$. Notice
that for prime $N$ \eqs{susiete} reduces to \eqs{susi}.

\subsection{Flux sums} 
%
%

The basic sums we have to consider are of the form 
\begin{equation} I(m,N)=
\sum_{\lambda=0}^{N-1} \ex^{\frac{i\pi m}{N} \lambda (N-\lambda)}=
\sum_{\lambda=0}^{N-1} \ex^{i\pi\frac{N-1}{N}m \lambda^2},
\la{zergs}
\end{equation} for $1\leq m \leq N-1$, and discrete Fourier
transformations thereof
\begin{equation}
\sum_{\lambda=0}^{N-1} \ex^{\pm\frac{2 i\pi}{N}
\lambda\lambda'}\ex^{i\pi\frac{N-1}{N}m \lambda^2},
\la{protoss}
\end{equation} from which the sums over fluxes can be easily
computed.  The basic sum \eqs{zergs} is related to a standard
Gauss sum
$G(m,N)=\sum_{r~ \text{\scriptsize mod $N$}} \ex^{2i\pi m r^2/N}$
\cite{numbertheory}. In fact, $I(m,N)=I(m+N,N)$ and, since $N$ is
odd, it suffices to consider the case where $m$ is even. But in
this case 
\begin{equation} I(2a,N)=\sum_{\lambda=0}^{N-1}
\ex^{i\pi\frac{N-1}{N}2a\lambda^2}= 
\sum_{\lambda} \ex^{-2i\pi a\lambda^2/N}=\overline{G(a,N)}.
\la{mutalisk}
\end{equation} 
Now, when $a=1$, 
\begin{equation}
G(1,N)=\frac{\sqrt{N}}{2}(1+i)\left(1+\ex^{-\frac{i\pi
N}{2}}\right),
\la{Gaussum}
\end{equation} (\cite{numbertheory}, p. $165$.) Moreover, for
$a>1$ and $N$ an {\sl odd prime}, 
\begin{equation}
G(a,N)=\left(\frac{a}{N}\right)\,G(1,N),
\la{Gaussum2}
\end{equation}
where $\left(\frac{a}{N}\right)$ is the Legendre symbol
\cite{numbertheory}, which is $+1$ if $a$ is a perfect square
($\mod~N$) and $-1$ otherwise. Hence, taking
\eqs{mutalisk}-\eqs{Gaussum2} into account we have the result
\begin{equation}
\sum_{\lambda=0}^{N-1} \ex^{i\pi\frac{N-1}{N}m \lambda^2}
=\epsilon(m)\sqrt{N} \ex^{-\frac{i\pi}{8}(N-1)^2},
\la{supinador}
\end{equation}
where

\begin{equation}
\epsilon(m)=\begin{cases}
\left(\frac{m/2}{N}\right), & \text{$m$
even},\\ \\
\left(\frac{(m+N)/2}{N}\right), & \text{$m$ odd},
\end{cases}
\la{sudiezseis}
\end{equation}

If $kh=-1~\mod~ N$, $\epsilon(k)=\epsilon(h)$ for $N=5 ~\mod~ 4$,
and 
$\epsilon(k)=-\epsilon(h)$ for $N=3 ~\mod ~4$. This property is
essential in proving the second relation in \eqs{sucuatro}. 

We also have the identity

\begin{equation}
\sum_{\lambda=0}^{N-1} \ex^{\pm \frac{2 i\pi}{N}
\lambda\lambda'}=N\delta_{\lambda',0},
\la{prono}
\end{equation}

and the fundamental result

\begin{equation}
\sum_{\lambda=0}^{N-1} \ex^{\pm\frac{2 i\pi}{N}
\lambda\lambda'}\ex^{i\pi\frac{N-1}{N}m \lambda^2}=
\epsilon(m)\sqrt{N}\ex^{-\frac{i\pi}{8}(N-1)^2}
\ex^{i\pi\frac{N-1}{N}h(\lambda')^2},
\la{pronador}
\end{equation}
with $mh=-1~\mod~N$ and $N$ an {\sl odd prime}. 

Now, given \eqs{supinador}, the basic sum over fluxes 
$\sum_{v} \ex^{i\pi\frac{N-1}{N}m v^2}$ can be computed in terms
of
\eqs{zergs} as follows -- see \cite{vw}, eq. (3.21)-(3.22):
\begin{equation}
\sum_{v\in H^{2}(X,\IZ_N)} \ex^{i\pi\frac{N-1}{N}m v^2} =
I(m,N)^{b_2^{+}} \,\overline {I(m,N)}{\,}^{b_2^{-}},
\la{terran}
\end{equation}
so one has (for {\it prime} $N$)
\begin{equation}
\sum_{v\in H^{2}(X,\IZ_N)} \ex^{i\pi\frac{N-1}{N}m v^2} =
\big(\epsilon(m)\big)^{b_2}N^{b_2/2}\ex^{-\frac{i\pi}{8}(N-1)^2
\sigma},
\la{terran2}
\end{equation}
and also, from \eqs{prono} and \eqs{pronador}  
\begin{equation}
\sum_{v\in H^{2}(X,\IZ_N)} \ex^{ \frac{2 i\pi}{N} u\cdot
v}=N^{b_2}\delta_{u,0},
\la{terran3}
\end{equation}

\begin{equation}
\sum_{v\in H^{2}(X,\IZ_N)} \ex^{\frac{2 i\pi}{N} u\cdot
v}\ex^{i\pi\frac{N-1}{N}m v^2}=
\big(\epsilon(m)\big)^{b_2}N^{b_2/2}\ex^{-\frac{i\pi}{8}(N-1)^2
\sigma}
\ex^{i\pi\frac{N-1}{N}hu^2},
\la{terran4}
\end{equation}
with $mh=-1~\mod~N$ as above.

\vskip1cm

\subsection{$SU(N)$ characters}
We have seen above that the corrections to the $SU(N)$ partition
function near the canonical  divisor of the four-manifold $X$ are
given in terms  of the level one characters $\chi_\lambda$ of the
$SU(N)$ WZW model. These are defined as \cite{izzy}
\begin{equation}
\chi_\lambda (\tau)=\frac{1}{\eta(\tau)^{N-1}}\sum_{\vec w\in
[\lambda]}
\ex^{i\pi\tau \vec w^2},\qquad \lambda\in\IZ~\mod~N,
\la{chars}
\end{equation}
where $[\lambda]$ is the $\lambda$-th conjugacy class of $SU(N)$,
and the identification $\chi_\lambda (\tau)=\chi_{\lambda+N}
(\tau)$ is understood. Also, from the symmetry properties of the inverse
Cartan matrix \eqs{cartan} it follows that $\chi_\lambda =
\chi_{N-\lambda}$. 
$\lambda\,$=$\,0~\mod~N$  corresponds to $\vec w$ in the root
lattice, while for 
$1\leq\lambda\leq N-1$,    
$[\lambda]=\{\vec w\in \Lambda_{\mbox{\rm\tiny weight}}:\vec
w=\vec\alpha^\lambda+\sum_{n^{\lambda'}\in\IZ}n^{\lambda'}
\vec\alpha_{\lambda'}\}$. $\vec\alpha_{\lambda}$ are the simple
roots and 
$\vec\alpha^\lambda$ the fundamental weights of $SU(N)$,
normalized in such a way that the inverse Cartan matrix
$A^{\lambda\lambda'}$ has the standard form 
\begin{equation} 
A^{\lambda\lambda'}=\vec\alpha^\lambda\cdot\vec\alpha^{\lambda'}=
{\rm Inf}~\{\lambda,\lambda'\}-\frac{\lambda\lambda'}{N},\qquad
1\leq\lambda,\lambda'\leq N-1.
\la{cartan}
\end{equation} 
The characters \eqs{chars} have the following properties under the
modular  group \cite{izzy}
\bea 
\chi_\lambda (\tau+1)&=&
\ex^{-\frac{i\pi}{12}(N-1)}\ex^{i\pi\frac{N-1}{N}\lambda^2}
\chi_\lambda (\tau),\ret
\chi_\lambda (-1/\tau)&=&\frac{1}{\sqrt{N}}\sum_{\lambda'=0}^{N-1} 
\ex^{-\frac{2 i\pi}{N}\lambda\lambda'}\chi_{\lambda'}(\tau).\ret
\la{sudoce}
\eea

From the characters $\chi_\lambda$ we introduce the linear
combinations  ($N>2$ and prime)
\begin{equation}
\chi_{m,\lambda}(\tau)=\frac{1}{N} \sum_{\lambda'=0}^{N-1}
\ex^{-\frac{2 i\pi}{N}
\lambda\lambda'}\ex^{i\pi\frac{N-1}{N}m (\lambda')^2}
\chi_{\lambda'}(\tau),\quad 0\leq m,\lambda\leq N-1, 
\label{suonce}
\end{equation} 
which have the ciclicity property
$\chi_{m+N,\lambda}=\chi_{m,\lambda}=\chi_{m,\lambda+N}$ since
$N$ is odd. Under the modular group one has
\bea 
\chi_{m,\lambda} (\tau+1)\!\!\! &=& \!\!
\ex^{-\frac{i\pi}{12}(N-1)}
\chi_ {m+1,\lambda}(\tau),\ret
\chi_{0,\lambda} (-1/\tau)\!\!\! &=& \!\!
\frac{1}{\sqrt{N}} \chi_{\lambda}(\tau),\ret
\chi_{m,\lambda} (-1/\tau)\!\!\! &=&
\!\!\epsilon(m)\,\ex^{-\frac{i\pi}{8}(N-1)^2}
\ex^{i\pi\frac{N-1}{N}h\lambda^2}\chi_{m,h\lambda} (\tau),\quad
m>0,
\ret
\la{suquince}
\eea
with $mh=-1~\mod~ N$.

\vfill
\newpage



\begin{thebibliography}{99}

\def\np{Nucl. Phys.}
\def\pl{Phys. Lett.} 
\def\pre{Phys. Rep.} 
\def\prl{Phys. Rev. Lett.}
\def\pr{Phys. Rev.} 
\def\ap{Ann. Phys.} 
\def\cmp{Commun. Math. Phys.}
\def\ijmp{Int. J. Mod. Phys.} 
\def\mpl{Mod. Phys. Lett.} 
\def\lmp{Lett. Math. Phys.} 
\def\bams{Bull. AMS} 
\def\am{Ann. of Math.} 
\def\jpsc{J. Phys. Soc. Jap.} 
\def\ijm{Int. J. Math.}
\def\jmp{J. Math. Phys.} 
\def\jgp{J. Geom. Phys.} 
\def\jdg{J. Diff. Geom.}
\def\plms{Proc. London Math. Soc.}
\def\mrl{Math. Res. Lett.}
\def\inma{Invent. Math.}
\def\tam{Trans. Am. Math. Soc.}
\def\jhep{J. High Energy Phys.}
\def\atmp{Adv. Theor. Math. Phys.}
\def\topo{Topology}


\bibitem{vw} C. Vafa and E. Witten, ``A strong coupling test of
$S$-duality," {\tt hep-th/9408074}; {\sl Nucl. Phys.} {\bf B431}
(1994) 3.

\bibitem{yamron} J. P. Yamron, ``Topological actions in twisted
supersymmetric theories," {\sl Phys. Lett.} {\bf B213} (1988)
325. 

\bibitem{ene4} J.M.F. Labastida and Carlos Lozano,
``Mathai-Quillen formulation  of twisted $\cn=4$ supersymmetric
gauge theories in four  dimensions," {\sl Nucl. Phys.} {\bf B502}
(1997) 741; {\tt hep-th/9702106}.

\bibitem{monoli} C. Montonen and D. Olive, ``Magnetic monopoles
as gauge  particles?" {\sl Phys. Lett.} {\bf B72} (1977) 117.

\bibitem{gthooft} G.'t Hooft, ``On the phase transition  towards
permanent quark  confinement," {\sl Nucl. Phys.} {\bf B138}
(1978) 1;  ``A property of electric and magnetic flux in
non-Abelian gauge theories,"  {\sl Nucl. Phys.} {\bf B153} (1979)
141.

\bibitem{italia} L. Girardello, A. Giveon, M. Porrati and A.
Zaffaroni, ``S-duality in $\cn$=4 Yang-Mills theories with
general gauge groups,"  {\sl Nucl.Phys.} {\bf B448} (1995) 127;
{\tt hep-th/9502057}.

\bibitem{htwist} J.M.F. Labastida and Carlos Lozano, ``Duality in
twisted ${\cal N}= 4$ supersymmetric gauge theories in four
dimensions," {\sl Nucl. Phys.}  {\bf B537} (1999) 203; {\tt
hep-th/9806032}.

\bibitem{estrings} J.A. Minahan, D. Nemeschansky, C. Vafa and
N.P. Warner,  ``E-strings and $\cn=4$ topological Yang-Mills
theories," {\sl  Nucl. Phys.} {\bf B527} (1998) 581; {\tt
hep-th/9802168}.

\bibitem{faro} J.M.F. Labastida and Carlos Lozano,   ``Duality 
in the context of topological quantum field theory," {\tt
hep-th/9901161}.

\bibitem{wijmp} E. Witten, ``Supersymmetric Yang-Mills theory on
a four-manifold,"   {\sl J. Math. Phys.}
{\bf 35} (1994) 5101; {\tt hep-th/9403193}.

\bibitem{tqft} E. Witten, ``Topological quantum field theory,"
{\sl Commun. Math. Phys.} {\bf 117} (1988) 353.

\bibitem{marcus}N. Marcus, ``The other topological twisting of
$\cn=4$  Yang-Mills," {\sl Nucl. Phys.} {\bf B452} (1995) 331;
{\tt hep-th/9506002}.

\bibitem{blauthomp} M. Blau and G. Thompson, ``Aspects of
$N_{T}\geq 2$ topological gauge theories and D-branes," {\sl
Nucl. Phys.} {\bf B492}  (1997) 545; {\tt hep-th/9612143}.

\bibitem{moorewitten} G. Moore and E. Witten, ``Integration over
the $u$-plane in Donaldson theory," {\sl Adv. Theor. Math. Phys.} 
{\bf 1} (1998) 298; {\tt hep-th/9709193}.

\bibitem{masas} J.M.F. Labastida and Carlos Lozano, ``Mass
perturbations  in twisted $\cn=4$ supersymmetric gauge theories,"
{\sl Nucl. Phys.}  {\bf B518} (1998) 37; {\tt hep-th/9711132}.

\bibitem{coreatres} R. Dijkgraaf, J.-S. Park and B.J. Schroers, 
``$\cn=4$ supersymmetric Yang-Mills theory on a K{\"a}hler
surface",  {\tt hep-th/9801066}.

\bibitem{donagi} R. Donagi and E. Witten, ``Supersymmetric
Yang-Mills  theory and integrable systems," {\sl Nucl. Phys.}
{\bf B460} (1996) 299;  {\tt hep-th/9510101}.

\bibitem{apostol} T.M. Apostol, ``Modular functions and Dirichlet
series in number  theory," (Springer-Verlag, New York, 1976.)

\bibitem{polynom} J.M.F. Labastida and M. Mari\~no, ``Polynomial
invariants for $SU(2)$ monopoles," {\sl Nucl. Phys.}  {\bf B456}
(1995) 633; {\tt hep-th/9507140}.

\bibitem{monopole} E. Witten, ``Monopoles and four-manifolds," 
{\tt hep-th/9411102}; {\sl Math. Res. Letters} {\bf 1}  (1994)
769.

\bibitem{mmtwo} M. Mari\~no and G. Moore, ``The Donaldson-Witten
function for gauge groups of rank larger than one," {\sl Commun.
Math. Phys.} {\bf 199} (1998)  25; {\tt hep-th/9802185}.

\bibitem{yoshioka} K. Yoshioka, ``Betti numbers of moduli of
stable sheaves on some  surfaces," {\sl Nucl. Phys. B (Proc.
Suppl.)} {\bf 46} (1996) 263; ``Chamber structure of
polarizations and the moduli  of stable sheaves on a ruled
surface," {\sl Int. J. Math.} {\bf 7} (1996)  411.

\bibitem{dijkfvbr} R. Dijkgraaf, ``The mathematics of
fivebranes," {\tt hep-th/9810157}.

\bibitem{wiads2} E. Witten, ``AdS/CFT correspondence and
topological field theory," {\sl\jhep} {\bf 9812} (1998) 012; 
{\tt hep-th/9812012}.

\bibitem{malda} 
J. Maldacena, ``The large N limit of
superconformal field theories  and supergravity," {\sl Adv.
Theor. Math. Phys.} {\bf 2} (1998) 231,  {\tt hep-th/9711200}; 
\hfil\\
S.S. Gubser, I. Klebanov and A.M. Polyakov,
``Gauge theory correlators  from non-critical string theory,"
{\sl Phys. Lett.} {\bf B428} (1998) 105, {\tt hep-th/9802109};
\hfil\\
 E. Witten, ``Anti-de Sitter space and
holography,"  {\sl Adv. Theor. Math. Phys.} {\bf 2} (1998) 253;
{\tt hep-th/9802150}.

\bibitem{gopaku} 
R. Gopakumar and C. Vafa, ``Topological gravity
as large N  topological gauge theory,"   {\sl Adv. Theor. Math.
Phys.} {\bf 2} (1998) 413, {\tt hep-th/9802016}; ``On the gauge
theory/geometry  correspondence," {\tt hep-th/9811131}.

\bibitem{hull}
C.M. Hull, ``Timelike T-duality, de Sitter space, large $N$ gauge 
theories and topological field theory," 
{\sl J. High Energy Phys.} {\bf 9807} (1998) 021; {\tt hep-th/9806146}.

\bibitem{numbertheory}
Hua Loo Keng, ``Introduction to number theory," (Springer-Verlag, Berlin,
1982.)

\bibitem{izzy} C. Itzykson, ``Level one Kac-Moody characters and
modular invariants," {\sl Nucl. Phys. B (Proc. Suppl.)} {\bf 5}
(1988) 150.


\end{thebibliography}
\end{document}